\documentclass[aps,prc,reprint,superscriptaddress,amsmath,amssymb,showpacs]{revtex4-1}

\usepackage{graphicx}
\usepackage{bm}
\usepackage{multirow}

\begin{document}


\title{Total Absorption Spectroscopy Study of the Beta Decay of $^{86}$Br and $^{91}$Rb}



\author{S.~Rice}
\affiliation{University of Surrey, Department of Physics, Guildford GU2 7XH, United Kingdom}
\author{A.~Algora}
\email[Corresponding author:]{algora@ific.uv.es}
\affiliation{Instituto de Fisica Corpuscular (CSIC-Universitat de Valencia), Apdo. Correos 22085, E-46071 Valencia, Spain}
\affiliation{Institute of Nuclear Research, Debrecen, Hungary}
\author{J.~L.~Tain}
\author{E.~Valencia}
\author{J.~Agramunt}
\author{B.~Rubio}
\affiliation{Instituto de Fisica Corpuscular (CSIC-Universitat de Valencia),
  Apdo. Correos 22085, E-46071 Valencia, Spain}
\author{W.~Gelletly}
\affiliation{University of Surrey, Department of Physics, Guildford GU2 7XH, United Kingdom}
\affiliation{Instituto de Fisica Corpuscular (CSIC-Universitat de Valencia), Apdo. Correos 22085, E-46071 Valencia, Spain}
\author{P.H.~Regan}
\affiliation{University of Surrey, Department of Physics, Guildford GU2 7XH, United Kingdom}
\affiliation{National Physical Laboratory, Teddington, TW11 0LW, United Kingdom}
\author{A.-A.~Zakari-Issoufou}
\author{M.~Fallot}
\author{A.~Porta}
\affiliation{SUBATECH, CNRS/IN2P3, Universit\'e de Nantes, Ecole des Mines, F-44307 Nantes, France}

\author{J.~Rissanen}
\author{T.~Eronen}
\affiliation{University of Jyv\"askyl\"a, Department of Physics, P.O. Box 35, FI-40014 Jyv\"askyl\"a, Finland}

\author{J.~\"Ayst\"o}
\affiliation{Helsinki Institute of Physics, FI-00014 University of Helsinki, Finland}

\author{L.~Batist}
\affiliation{Petersburg Nuclear Physics Institute, RU-188300 Gatchina, Russia}

\author{M.~Bowry}
\affiliation{University of Surrey, Department of Physics, Guildford GU2 7XH, United Kingdom}

\author{V.~M.~Bui}
\affiliation{SUBATECH, CNRS/IN2P3, Universit\'e de Nantes, Ecole des Mines, F-44307 Nantes, France}

\author{R.~Caballero-Folch}
\affiliation{Universitat Politecnica de Catalunya, E-08028 Barcelona, Spain}

\author{D.~Cano-Ott}
\affiliation{Centro de Investigaciones Energ\'eticas Medioambientales y
  Tecn\'ologicas, E-28040 Madrid, Spain}

\author{V.-V.~Elomaa}
\affiliation{University of Jyv\"askyl\"a, Department of Physics, P.O. Box 35, FI-40014 Jyv\"askyl\"a, Finland}

\author{E.~Estevez}
\affiliation{Instituto de Fisica Corpuscular (CSIC-Universitat de Valencia),
  Apdo. Correos 22085, E-46071 Valencia, Spain}

\author{G.~F.~Farrelly}
\affiliation{University of Surrey, Department of Physics, Guildford GU2 7XH, United Kingdom}

\author{A.~R.~Garcia}
\affiliation{Centro de Investigaciones Energ\'eticas Medioambientales y
  Tecn\'ologicas, E-28040 Madrid, Spain}

\author{B.~Gomez-Hornillos}
\author{V.~Gorlychev}
\affiliation{Universitat Politecnica de Catalunya, E-08028 Barcelona, Spain}

\author{J.~Hakala}
\affiliation{University of Jyv\"askyl\"a, Department of Physics, P.O. Box 35, FI-40014 Jyv\"askyl\"a, Finland}

\author{M.~D.~Jordan}
\affiliation{Instituto de Fisica Corpuscular (CSIC-Universitat de Valencia),
  Apdo. Correos 22085, E-46071 Valencia, Spain}

\author{A.~Jokinen}
\author{V. S. Kolhinen}
\affiliation{University of Jyv\"askyl\"a, Department of Physics, P.O. Box 35, FI-40014 Jyv\"askyl\"a, Finland}

\author{F.~G.~Kondev}
\affiliation{Nuclear Engineering Division, Argonne National Laboratory, 
Argonne, Illinois 60439, USA}

\author{T.~Mart\'{\i}nez}
\affiliation{Centro de Investigaciones Energ\'eticas Medioambientales y
  Tecn\'ologicas, E-28040 Madrid, Spain}

\author{P.~Mason}
\affiliation{University of Surrey, Department of Physics, Guildford GU2 7XH, United Kingdom}

\author{E.~Mendoza}
\affiliation{Centro de Investigaciones Energ\'eticas Medioambientales y
  Tecn\'ologicas, E-28040 Madrid, Spain}

\author{I.~Moore}
\author{H.~Penttil\"a}
\affiliation{University of Jyv\"askyl\"a, Department of Physics, P.O. Box 35, FI-40014 Jyv\"askyl\"a, Finland}

\author{Zs.~Podoly\'ak}
\affiliation{University of Surrey, Department of Physics, Guildford GU2 7XH, United Kingdom}

\author{M.~Reponen}
\author{V.~Sonnenschein}
\affiliation{University of Jyv\"askyl\"a, Department of Physics, P.O. Box 35, FI-40014 Jyv\"askyl\"a, Finland}

\author{A.~A.~Sonzogni}
\affiliation{NNDC, Brookhaven National Laboratory, Upton, New York 11973, USA}

\author{P.~Sarriguren}
\affiliation{Instituto Estructura de la Materia, IEM-CSIC, Serrano 123, E-28006 Madrid, Spain }


\date{\today}

\begin{abstract}
The beta decays of $^{86}$Br and $^{91}$Rb  have been studied using the total absorption spectroscopy technique. The radioactive nuclei were produced at the IGISOL facility in Jyv\"askyl\"a and further purified using the JYFLTRAP. $^{86}$Br and $^{91}$Rb are considered to be major contributors to the decay heat in reactors. In addition $^{91}$Rb was used as a normalization point in direct measurements of mean gamma energies released in the beta decay of fission products by Rudstam {\it et al.} assuming that this decay was well known from high-resolution measurements. Our results show that both decays were suffering from the {\it Pandemonium} effect and that the results of Rudstam {\it et al.} should be renormalized. The relative impact of the studied decays in the prediction of the decay heat and antineutrino spectrum from reactors has been evaluated.

\end{abstract}



\maketitle

Beta decay studies can provide relevant information for fundamental physics, nuclear structure  and practical applications. One important application is in nuclear technology, where beta decay data are used for the evaluation of $\gamma$-ray and $\beta$ spectra emitted by fission products in a working reactor, after reactor shut down, in the nuclear waste generated and for the prediction of the spectrum of antineutrinos emitted by a reactor \cite{alg10,Fallot}. 

In recent years the summation calculation method has been the most widely used technique for the evaluation of the $\beta$- and $\gamma$- energy released from the fission products in a reactor or in the nuclear waste. The  inputs needed for these calculations are the mean- $\gamma$ and $\beta$ energies released in the beta decay of each fission product. The mean energies can be obtained from direct measurements of the gamma \cite{Rudstam90}  and the beta \cite{Tengblad} radiation emitted in each radioactive decay or can be deduced from evaluated nuclear data available in databases \cite{ensdf}. Most of the data, which are available in databases, come from measurements using conventional high-resolution gamma-ray spectroscopy, that can suffer from a systematic error known as the {\it Pandemonium} effect~\cite{har77}. This systematic error arises from the difficulty of detecting weak $\gamma$-ray cascades and (or) high-energy $\gamma$-rays because of the limited efficiency of the germanium detectors that are usually employed in conventional $\beta$-decay studies. As a result, the decay scheme deduced may be incomplete, and the beta decay probability distribution, deduced from the gamma intensity balance populating and de-exciting each level, may be incorrect. In practical terms this means erroneously assigning more beta intensity to lower-lying levels and this as a consequence leads to an overestimation of the mean beta energies and an underestimation of the mean gamma energies. 

To avoid this systematic error, the total absorption gamma-ray spectroscopy technique (TAGS) can be used. The technique aims at detecting gamma cascades rather than individual $\gamma$ rays using large $4\pi$ scintillation detectors. The advantage of this method over high-resolution germanium spectroscopy
to locate missing $\beta$ intensity has been demonstrated before, for cases measured using both techniques and including some measured with a highly efficient Ge array ~\cite{alg99,hu99,alg03}. 

In this article we present the results of measurements performed for two decays, $^{86}$Br and $^{91}$Rb, 
which are considered to be high priority contributors to the decay heat in reactors \cite{Nichols,Yoshida,Gupta2010}. Previous results from the same experimental campaign have already been  published  ~\cite{Zak15,Tain15}.  The total absorption measurement of the decay of $^{91}$Rb is of particular interest, since it was used as a calibration point for the mean gamma energy measurements reported by Rudstam {\it et al.} \cite{Rudstam90}, which are still widely used as a reference.  In the measurements of Rudstam {\it et al.}, a well collimated NaI(Tl) scintillation detector was used to detect single $\gamma$-rays from decay cascades of the mass separated fission products. From the measured spectrum a $\gamma$-ray intensity distribution was obtained after deconvolution with the measured spectrometer response. To derive the mean $\gamma$ energy from this distribution the intensity must be calibrated on an absolute scale. For this, the number of decays was obtained from selected transitions whose intensity was regarded as well known and were detected in an auxiliary Ge(Li) detector. To calibrate the absolute efficiency of the setup $^{91}$Rb was selected because it  has a relatively large $Q_{\beta} =5907(9)$~keV value \cite{Wang12} and the decay level scheme was regarded as being free from {\it Pandemonium}. Thus the calibration of the mean gamma energies in Ref.~\cite{Rudstam90} was done using an intensity  of 8.3(4)\% for the 345~keV  transition in $^{91}$Sr and matching the mean energy of the $^{91}$Rb distribution to  the high resolution value of 2335(33)~keV.
$^{91}$Rb was also measured by Greenwood {\it et al.} \cite{Greenwood} using the total absorption technique, but employing different analysis techniques. The present measurement will allow us to compare our data with Greenwood's results to further validate the measurements and the analysis techniques. 

The determination of the beta decay probability distribution free from the {\it Pandemonium} effect also makes it possible to compare the deduced strength with theoretical calculations. $^{91}$Rb lies in a transitional region characterized by shape changes \cite{Guz10}. For that reason it is also worth exploring the possibility of  infering its ground state shape from a comparison of the deduced beta strength in the daughter with theoretical calculations as was already performed for nuclei in the A$\sim$80 and A$\sim$190 regions \cite{Nacher,Poirier,Perez,Briz,Aguado}.  

$^{86}$Br decay is also of particular interest from the perspective of total absorption measurements. It has a large $Q_{\beta} =7633(3)$~keV value \cite{Wang12}, and the high resolution decay scheme is poorly known. Only 17 excited levels have been placed in $^{86}$Kr while the total number of levels expected to be fed, from level density considerations, is around 300. Thus one could expect a relatively large {\it Pandemonium} effect. This and the large contribution of this decay at cooling times around 100~s are the reasons to include this nucleus with high priority in the lists  \cite{Yoshida,Gupta2010} for decay heat data measurements using the TAGS technique. $^{86}$Br decay has also been considered recently in the framework of studying the pygmy dipole resonance (PDR) through beta decay \cite{Scheck16}. From this perspective, decays that 
preferentially populate $1^{-}$ levels that can be associated with the PDR inside a large $Q$ value of the decay are of particular interest. In Ref. \cite{Scheck16} it was concluded that in particular cases, beta decay populates levels associated with the PDR, but only a fraction of those, and this can be considered as a source of  complementary information for PDR studies. For this new application the TAGS technique is a source of reliable data on absolute intensities of beta decay transitions and on the decay branching ratios of the populated levels. In addition, from the comparison with the calculated $\beta$-strength distributions, information on the structure of these levels can be obtained. 

\begin{table}[h]
\caption{\label{Density_parameters} Level Density parameters used in the analysis for daughter isotopes (parameters given for the Gilbert-Cameron (GC) formulation \cite{Gilbert}, which is a combination of the Back Shifted Fermi Gas (BSFG) model \cite{BSFG_Dilg} plus the Constant Temperature (CT) model \cite{CT} for high excitation energy). The parameters are: the ground state position $\Delta$, the level density $a$ (for BSFG), nuclear temperature $T$ and the back-shift $E_0$ (for CT) and the matching point $E_x$ of the BSFG and CT models for the Gilbert and Cameron model.}
\begin{ruledtabular}
\begin{tabular}{|c|c|c|c|c|c|} 
&\multicolumn{5}{c|}{Level density Parameters}\\
Isotope&a&$\Delta$&T&E0&Ex\\ 
           &[1/MeV]&[MeV]&[MeV] &[MeV] &[MeV] \\ \hline
$^{86}$Kr&8.434&1.599&0.833&1.518&4.342\\
$^{91}$Sr&9.754&0.264&0.662&0.425&1.946\\ 
\end{tabular}
\end{ruledtabular}
\end{table}

\begin{table*}[h]
\caption{\label{Strength_parameters} Gamma strength function parameters  used in the analysis for daughter isotopes.}
\begin{ruledtabular}
\begin{tabular}{|c|c|c|c|c|c|c|c|c|c|} 
&\multicolumn{9}{c|}{Strength Function Parameters}\\
&\multicolumn{3}{c}{E1}&\multicolumn{3}{c}{M1}&\multicolumn{3}{c|}{E2}\\\cline{2-10}
Isotope&Energy&Width&$\sigma$&Energy&Width&$\sigma$&Energy&Width&$\sigma$\\
&[MeV]&[MeV]&[mb]&[MeV]&[MeV]&[mb]&[MeV]&[MeV]&[mb] \\ \hline
\multirow{2}{*}{$^{86}$Kr}&16.29&5.37&178.7&\multirow{2}{*}{9.30}&\multirow{2}{*}{4.00}&\multirow{2}{*}{19.67}&\multirow{2}{*}{14.29}&\multirow{2}{*}{5.08}&\multirow{2}{*}{1.78}\\
&17.17&5.94&161.63&&&&&&\\ \hline
\multirow{2}{*}{$^{91}$Sr}&16.08&5.24&193.81&\multirow{2}{*}{9.13}&\multirow{2}{*}{4.00}&\multirow{2}{*}{2.66}&\multirow{2}{*}{14.03}&\multirow{2}{*}{5.02}&\multirow{2}{*}{1.89}\\
&16.95&5.79&175.32&&&&&& 
\end{tabular}
\end{ruledtabular}
\end{table*}

\section*{The Experiment}

The measurements were performed at the  IGISOL facility ~\cite{ays01} of the University of Jyv\"askyl\"a as part of an experimental campaign aimed at measuring beta decays of nuclei that are important contributors to the decay heat and to the antineutrino spectrum in reactors. As already discussed in ~\cite{Zak15,Tain15}, the isotopes of interest were produced by proton-induced fission of uranium and first mass separated using the moderate resolution mass separator of IGISOL with a mass resolving power of approximately 500. Since the purity of the samples is of great importance for the measurements, the radioactive beam of the selected mass was further purified isotopically using the JYFLTRAP Penning trap~\cite{kol04,ero12}. Then, the extracted radioactive beam of the isotope of interest  was implanted at the centre of the total absorption spectrometer onto a tape which was moved periodically to reduce the impact of the daughter contamination in the measurements. The measurement cycles were selected according to the half-lives of the decays of interest.  
Behind the tape, at approximately 5 mm from the implantation point,  a 0.5~mm thick Si detector with a $\beta$-detection efficiency of about 25\% was placed. The implantation point was surrounded by the 
Valencia-Surrey Total Absorption Spectrometer {\it Rocinante}. This spectrometer is a cylindrical 12-fold
segmented BaF$_{2}$ detector with a length and external diameter of 25~cm, 
and a longitudinal hole of 5~cm diameter.
The separation between crystals in this spectrometer is provided by a thin optical reflector.
The total efficiency of the setup for detecting a single $\gamma$ ray is larger than 80\% (up to 10 MeV). Since the BaF$_{2}$ has an intrinsic background, coincidences with the beta detector were used to generate  $\beta$-gated TAGS spectra in the present analysis. Using coincidences also avoids the contribution of normal ambient background in the measured spectra. 

\section*{Analysis}

The first step in the analysis of the total absorption experiments is to determine the contaminants in the spectra to be analyzed. As mentioned earlier, the use of the beta-coincidence conditions, cleans the spectrum  of internal and ambient backgrounds, but daughter decay contamination and pulse pileup contributions have to be determined. Since we are dealing with a segmented detector, apart from the electronic pulse pile-up that affects a single detector module~\cite{cano1999pulse}, one must also consider the summing of signals from different detector modules \cite{Tain15}. To address this problem a new Monte Carlo (MC) procedure to determine their combined contribution has been implemented. The method is based on the random superposition of two of the stored events within the analog-to-digital converter (ADC) gate length. The normalization of the resulting summing-pileup spectrum is then calculated by the event rate and the ADC gate length as in Ref.~\cite{cano1999pulse}. 
Once the contributions of the contaminants have been determined, one can apply the analysis methods to the measured spectrum to obtain the feeding distribution. In this work as in earlier studies, we follow the procedures developed by the Valencia group~\cite{tai07a,tai07b}. 
 
For that we need to solve the TAGS inverse problem:

\begin{equation}\label{drf}
  d_i = \displaystyle\sum_{j=0}^{j_{max}}R_{ij}(B)f_j + C_i
\end{equation}

where $d_i$ is the content of bin $i$ in the  measured TAGS spectrum, $R_{ij}$ is the response matrix of the TAGS setup and represents the probability that a decay that feeds level $j$ in the level scheme of the daughter nucleus gives a count in bin $i$ of the TAGS spectrum, $f_j$ is the beta feeding to the level $j$ and $C_i$ is the contribution of the contaminants to bin $i$ of the TAGS spectrum. The response matrix $R_{ij}$ depends on the TAGS setup and on the assumed level scheme of the daughter nucleus (branching ratio matrix $B$). To calculate the response matrix the $B$ matrix for the levels in the daughter nucleus has to be determined first. For that the level scheme of the daughter nucleus is divided into two regions, a low excitation part and a high excitation part. Conventionally the levels of the low excitation part and their gamma decay branchings are taken from high resolution measurements available in the literature, since it is assumed that the gamma branching ratios of these levels are well known. Above a certain energy, the cut-off  energy, a continuum of possible levels divided into 40 keV bins is assumed.  From this energy up to the  decay $Q$ value, the statistical model is used to generate a branching ratio matrix for the high excitation part of the level scheme. The statistical model is based on a level density function and gamma strength functions of E1, M1, and E2 character. In the cases presented here, the parameters for the gamma strength function were taken from \cite{Capote2009} 
and the parameters of the level density function \cite{Gilbert,BSFG_Dilg,CT} were obtained from fits to the data available in \cite{Goriely2001,Demetriou2001,Capote2009}. Details of the parameters used are given in Tables~\ref{Density_parameters} and \ref{Strength_parameters}. As part of the optimisation procedure in the analysis, the cut-off energy and the parameters of the statistical model can be changed. Once the branching ratio matrix is defined, the $R_{ij}$ can be calculated recursively from responses previously determined using Monte Carlo simulations \cite{cano1999monte,cano1999pulse,agostinelli2003}. The Monte Carlo simulations were validated with measurements of the spectra of well known radioactive sources ($^{24}$Na, $^{60}$Co, $^{137}$Cs).  Once the $R$ response matrix is obtained, the Expectation Maximisation (EM) algorithm is applied to extract the beta feeding distributions from equation \ref{drf}. 

The feeding distributions obtained  from the analyses will then be used to calculate the mean gamma and beta energies released in the decay using the following relations: $ \overline{E_\gamma} = \sum_i E_i*I_i$, 
and  $ \overline{E_\beta} = \sum_i I_i*<E_\beta>_i$, where $E_i$ is the energy of the level $i$, $I_i$
is the normalized feeding to level $i$, and $<E_\beta>_i$ is the mean energy of the beta continuum populating level $i$. In the case of $^{91}$Rb decay, the normalized feeding distribution will also be used to deduce the beta strength for comparison with theoretical calculations. 

\section*{Decay of $^{91}$Rb}

\begin{figure}[ht]
 \begin{center}
 \includegraphics[width=8.6cm]{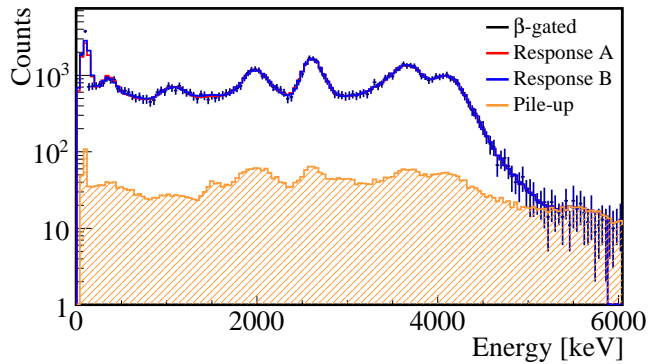}
 \caption{\label{91Rb_spectrum} (Color online) Relevant histograms for $^{91}$Rb decay: measured spectrum (dotted line), summing-pileup contribution (orange line),  reconstructed spectrum response A (red line), reconstructed spectrum response B (blue line) . Response A corresponds to the conventional analysis. Response B has additional optimization on the branching ratio matrix to reproduce the measured $\gamma$ intensities in high resolution experiments.}
\end{center}
\end{figure}

The tape cycle for the measurement of the decay of $^{91}$Rb was set to 174.8 s. With this measuring cycle the daughter decay contamination can be estimated to be approximately 0.1 \% from the solution of the Bateman equations using 58.2(3) s \cite{Baglin13} for the decay half-life of $^{91}$Rb, and 9.65(6) h for the half-life of the daughter $^{91}$Sr. For that reason the daughter activity was not measured separately. In this case the only contamination in the beta-gated spectrum is the summing-pileup, as showed in Fig. \ref{91Rb_spectrum}. 

For the analysis we need to define the branching ratio matrix of the daughter nucleus level scheme. As mentioned earlier this requires the combination of the known levels from high resolution measurements and complementing the missing information up to the Q value with the statistical model. According to the latest ENSDF evaluation \cite{Baglin13} the level scheme of the daughter nucleus is poorly known in terms of spin-parity assignments, since only one level in the daughter nucleus has a firm spin-parity assignment in the decay level scheme. The missing spins and parities of the levels needed to be estimated. For that purpose, the known gamma transitions between levels were used in combination with the expectation that most gamma transitions will occur via the most probable E1, E2 and M1 gamma-ray transitions, resulting in a range of options available for the missing spins and parities. A number of these levels are recorded to decay via E2/M1 transitions to the 94 keV (3/2$^+$) state, resulting in the initial decaying level  probably being  1/2$^+$, 3/2$^+$ or 5/2$^+$. In addition, the beta decay feeding distribution available in ENSDF was also used initially when postulating options for the spin-parity assignments. The large number of degrees of freedom now available via these options results in a range of level schemes.  These level schemes were considered up to different energy level thresholds for the application of the statistical model during the analysis.

\begin{figure}[ht]
 \begin{center}
 \includegraphics[width=8.6cm]{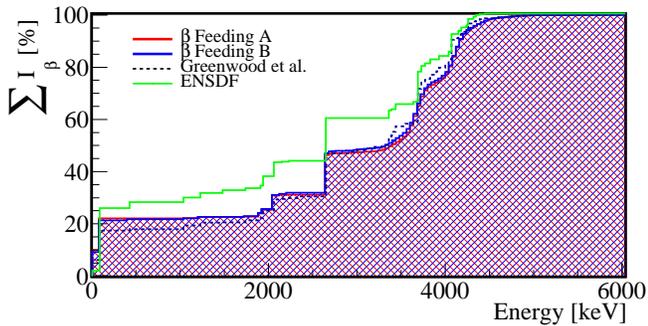}
 \caption{\label{91Rb_feeding} (Color online) Comparison of the accumulated feeding distributions obtained in this work for the decay of $^{91}$Rb with the distributions from earlier high resolution measurements \cite{Baglin13} and with that obtained by Greenwood {\it et al.} \cite{Greenwood}. }
\end{center}
\end{figure}

The parameters used in the final analysis for the level density parametrization and for the gamma strength functions are given in Tables \ref{Density_parameters} and \ref{Strength_parameters}. For the continuum part of the level scheme several possibilities were tested for the level density parametrization (Back Shifted Fermi Gas formula, Constant Temperature and a combination of both, the Gilbert Cameron formula  \cite{BSFG_Dilg,CT,Gilbert}).  Similar results were obtained in the analysis for the Gilbert-Cameron formula and for the Constant Temperature model. In many of the analyses performed it was found that low cut-off  energies in the known level scheme resulted  in a poor reproduction of the peak around 2600 keV in the total absorption spectrum. It is worth noting that the spin and parity of the parent $^{91}$Rb is 3/2$^{(-)}$ \cite{Baglin13}. For for that reason also analyses were performed assuming a 3/2$^+$ assignment, and RIBF accordingly considering  other ranges of populated states (allowed and first forbidden decays) than that in the case of the 3/2$^-$ ground state assumption. Those analyses provided a poorer reproduction of the data. As a result, in the final accepted analysis, we have assumed a cut-off energy at 2680 keV and allowed and first forbidden decays were considered assuming a parent state with 3/2$^-$.  The results of the final analyses are presented in Figs. \ref{91Rb_spectrum} and \ref{91Rb_feeding}. In Fig.  \ref{91Rb_spectrum} two analyses are provided. Analysis labelled A, represents the analysis performed conventionally. Analysis B, is an analysis performed using a slightly modified branching ratio matrix, in order to reproduce the experimental gamma intensities obtained in high-resolution experiments.  This optimization is performed by adjusting the gamma feeding from the levels in the continuum to the discrete levels in the branching ratio matrix of the accepted analysis (labelled A). 
The results presented in Figs. \ref{91Rb_spectrum} and \ref{91Rb_feeding} show that the quality of the reproduction of the measured decay spectrum is very similar for both forms of analysis. Small differences emerge in the feeding distribution, as can be seen in Fig. \ref{91Rb_feeding}, which appear mainly for levels that have direct gamma connections to the ground state. The analysis B is able to reproduce the gamma intensity de-exciting the level at 439 keV within 3 \%, which is relevant in this context because the gamma ray of 345.5 keV de-exiting this level, with an intensity error of 5 \%, was used as the global normalization point by Rudstam {\it et al.} in their mean gamma energy measurements. 

Both feeding distributions obtained are similar to the one obtained by Greenwood \cite{Greenwood}.  From the two distributions, the feeding distribution obtained with the optimized branching ratio matrix lies closer to the Greenwood result. The three total absorption results clearly differ from the ENSDF data \cite{Baglin13} based on high resolution measurements. From our conventional analysis a ground state feeding of 10.2 \% is obtained, which can be compared with the value of Greenwood {\it et al.} \cite{Greenwood2} of 6.2 \%, while the optimized branching ratio matrix result is slightly smaller at 9.2 \%.  Those values can be compared with the ENSDF adopted value of 2 (5) \% \cite{Baglin13}. But we must mention that the division of the feeding values between the ground state and first excited level at 93.4 keV should be taken with caution, since the two levels lie very close in energy as already presented in Greenwood {\it et al.} \cite{Greenwood}. As an additional test, we also performed an analysis fixing the ground state and first state feeding to the Greenwood values. In this last case the quality of the fit to the data was clearly much worse than the accepted ones. 

\begin{table}[h]	
\caption{\label{HeatRb} Mean average energy for $\beta$-particles and $\gamma$ rays (all collected photons) from the decay of $^{91}$Rb compared with the value included in the ENSDF database and with the values obtained by Greenwood {\it et al.}  and Rudstam {\it et al.}}
\begin{ruledtabular}
\begin{tabular}{|l|c|c|}
&$\bar{E_\gamma}$ [keV]&$\bar{E_\beta}$ [keV]\\\hline
Present result & 2669(95)&1389(44)\\
Greenwood {\it et al.} & 2708(76) &1367(44)\\
Rudstam {\it et al.} & 2335(33) &  1560(30) \\
ENSDF & 2342(45) & 1619(19) \\
\end{tabular}
\end{ruledtabular}
\end{table}

\begin{figure}[ht]
 \begin{center}
 \includegraphics[width=8.6cm]{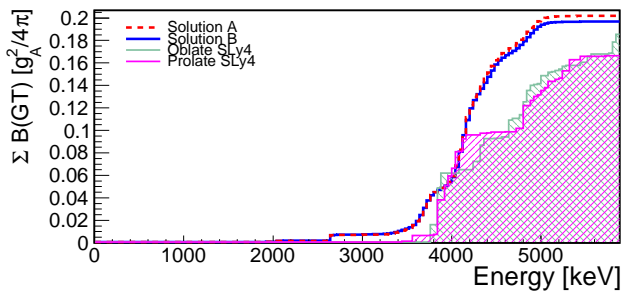}
 \caption{ \label{91Rb_strength} (Color online) Accumulated strength of the decay of $^{91}$Rb obtained for the two analyses presented in this work compared with QRPA calculations assuming oblate and prolate shapes for the ground state of $^{91}$Rb.}
\end{center}
\end{figure}

In Table \ref{HeatRb} we present a comparison of the deduced mean energies from the present work with the values determined from the Greenwood data and with the value used by Rudstam {\it et al.} In the table we quote the mean energies deduced from the results obtained from the optimized branching ratio matrix analysis (analysis B). The error in the mean energies is evaluated from the differences in the mean gamma and beta values obtained from several analyses, that provided a good description of the experimental data. The present value is close to the result of Greenwood and shows a large difference with the value used by Rudstam, which was based on earlier high resolution measurements. This result as well as the comparison presented in Fig. \ref{91Rb_feeding} confirm that the value used by Rudstam as a normalization point,  suffered from the {\it Pandemonium} effect. For that reason all mean gamma energies published in Rudstam {\it et al.} should be multiplied by 1.14.

\begin{figure}[ht]
 \begin{center}
 \includegraphics[width=8.6cm]{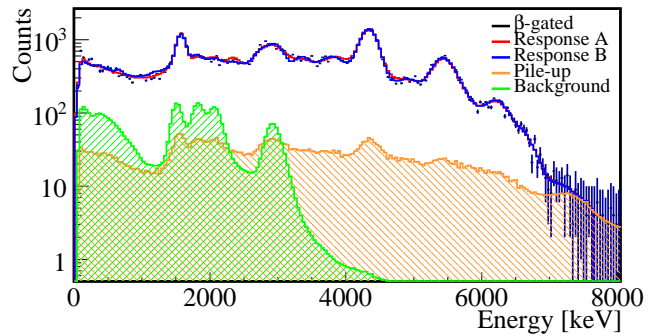}
 \caption{ \label{86Br_spectrum} (Color online) Relevant histograms for $^{86}$Br decay: measured spectrum (points with errors), reconstructed spectrum response A (blue line), reconstructed spectrum response B (red line) summing-pileup contribution (orange line),  background (green line).}
\end{center}
\end{figure}

As mentioned in the introduction $^{91}$Rb lies in a region of shape transitions. For that reason it is also worth examining how well the beta-decay strength of $^{91}$Rb is reproduced by theoretical calculations that assume different possible shapes for its ground state. The measured strength is compared in Fig. \ref{91Rb_strength} with results from deformed quasiparticle random-phase approximation (QRPA) calculations. In this formalism, a selfconsistent quasiparticle basis is first constructed from deformed Skyrme Hartree-Fock calculations with pairing correlations in the BCS approximation. 
Then, a separable spin-isospin residual interaction is included in both particle-hole and particle-particle channels and treated in the QRPA\cite{Sarriguren}.

The total energy as a function of the quadrupole deformation parameter shows two minima, one oblate at $\beta=-0.12$, which is the ground state, and another prolate at $\beta=0.10$ at about 300 keV excitation energy. The minima are very shallow with practically no barrier between them.



Figure \ref{91Rb_strength} shows the accumulated Gamow-Teller strength for the oblate and prolate shapes of $^{91}$Rb calculated in QRPA with the force SLy4. A standard quenching factor $(g_A/g_V)_{\rm eff} = 0.77 (g_A/g_V)$ is included in the calculations to compare with the data. In general, the agreement with experiment is very reasonable. There is basically no strength at low energy. The strength is concentrated at around 4 MeV and 5 MeV in the calculations. It is more fragmented and spread in the experiment, but again concentrated at about 4 MeV. The total strength contained in the $Q_{\beta}$ energy window is also comparable, although somewhat underestimated. It is also worth mentioning the similarity between the strength distributions of both oblate and prolate shapes that would prevent in this case the use of these experiments to determine deformation. The absence of GT strength observed in the calculations below 3-4 MeV is understood from the fact that the formalism deals only with allowed GT transitions. Indeed, the neutron states close to the neutron Fermi level are immersed in the group of states split from the spherical shells $g_{7/2}$ and $d_{5/2}$, which are positive parity states that cannot be connected with allowed transitions with the negative parity states coming from the $f_{5/2}$ and $p_{3/2}$ shells located in the vicinity of the proton Fermi level. Thus, most probably, the observed strength in the low-lying excitation energy has its origin in forbidden transitions involving a change in the parity of the states, which are not included in calculations in the present formalism.

\section*{Decay of $^{86}$Br}

The $\beta^-$ decay of $^{86}$Br proceeds to the stable nucleus $^{86}$Kr, therefore there is no daughter contamination for this decay. As in the $^{91}$Rb case, the pileup was calculated according to the recently developed procedure \cite{Tain15}. A preliminary analysis of the spectra cleaned of pileup  highlighted that there is a small amount of contamination in the beta-gated spectra. Since the production of the isotope was continuously checked and pure, the contamination was identified as a small background contribution, due to an increased level of noise in the silicon detector in one of the runs. Possible solutions to eliminate this contamination are the exclusion of the run from the analysis or to increase the threshold of the silicon detector, but since this run contained an important part of the statistics, we decided to use an alternative solution. 
In the analysis of this case we have subtracted from the beta-gated spectrum a background spectrum with beam-on, from which its own pileup had been previously subtracted. The level of subtraction was determined from a comparison with the clean run. The resulting spectra, with all the contributions are presented in Fig. \ref{86Br_spectrum}, where the results of the reconstructed spectra after the analyses are also shown.  

The first step in the deconvolution process is the determination of the branching ratio matrix. As discussed in the $^{91}$Rb case, the three statistical models (GC, BSFG and CT \cite{Gilbert,BSFG_Dilg,CT}) were fitted to the mixture of experimental and theoretical data to obtain the relevant level density parameters. Those  resulting from the GC model are summarised in Table~\ref{Density_parameters}. Also in Table~\ref{Strength_parameters} the gamma strength parameters  used in the construction of the branching ratio matrix for the daughter isotope  $^{86}$Kr are provided. 

The level scheme of the daughter $^{86}$Kr is better known than in the $^{91}$Sr case. Up to the level at an  excitation energy of 3099 keV, only two levels have uncertain spin-parity assignments. In addition, a recent ENSDF evaluation \cite{ensdf86} has included some new levels from a $^{86}$Kr$(n,n^{'})$$^{86}$Kr study from Fotiades {\it et al} \cite{Fotiades} and slightly revised the excitation energies of some levels compared with the earlier evaluation \cite{ensdf86_old}. 

An important change in the new evaluation of the decay of $^{86}$Br is the new spin-parity  assignment  of the ground state. Previously the spin-parity assignment of this state was $J^{\pi}=2^{-}$, based on the systematics from $^{82-84}$Br, but a relatively recent study by Porquet {\it et al.} \cite{Porquet09} suggested a possible $1^{-}$ asignment arising from the lowest energy state in the $\pi p_{3/2}\nu d_{5/2}$ multiplet. This new value has been assigned to the ground state in the new ENSDF evaluation \cite{ensdf86}. In our analyses both options were used, the $1^{-}$ cases providing better fits of the total absorption data, in particular to the region of the spectra around the peak at 2250 keV and in the region between 3500 and 4000 keV. 

\begin{figure}[ht]
 \begin{center}
 \includegraphics[width=8.6cm]{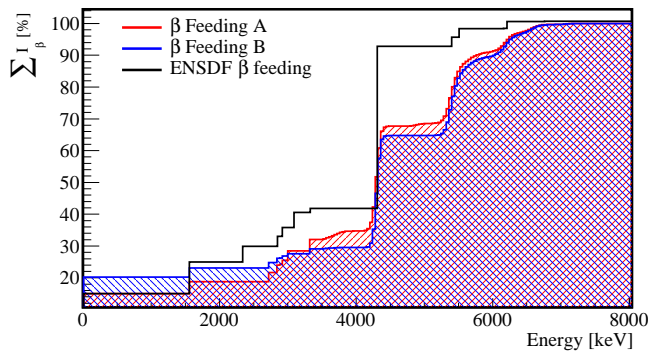}
 \caption{\label{86Br_feeding} (Color online) Comparison of the accumulated feeding distributions obtained in this work  for the decay of $^{86}$Br with the distributions from earlier high resolution measurements \cite{ensdf86}. }
\end{center}
\end{figure}

The final accepted analyses were performed using the $1^{-}$ assignment for the parent ground state and a cut-off energy in the known level scheme at 3560 keV. 
Allowed and first forbidden transitions were considered. The results of those analyses are presented in Figs. \ref{86Br_spectrum} and \ref{86Br_feeding}. 
As in the  $^{91}$Rb case, in Fig.  \ref{86Br_spectrum} two analyses are provided. Analysis labelled A, represents the analysis performed conventionally. Analysis B, is an analysis performed using a slightly modified branching ratio matrix, in order to reproduce the experimental gamma intensities obtained in high-resolution experiments. In this particular decay the result from the conventional analysis (labelled A) gave a larger discrepancy (41 \%) in the reproduction of the gamma intensity from the first excited state when compared with high resolution measurements. After the optimization of the branching ratio matrix, (analysis B), the gamma intensity de-exciting the first excited state is reproduced within 5 \%.  
The results presented in Figs. \ref{86Br_spectrum} and \ref{86Br_feeding} show that the quality of the reproduction of the measured decay spectrum is very similar for both analyses, being slightly worse for the adjusted one. Compared to the $^{91}$Rb case, slightly larger differences appear in the feeding distributions, as can be seen in Fig. \ref{86Br_feeding}, in particular analysis B, with the optimized branching ratio matrix, which provides a larger ground state feeding value.  As in the $^{91}$Rb case, the two total absorption results clearly differ from the ENSDF data \cite{ensdf86} based on high resolution measurements, which points to a decay suffering from the {\it Pandemonium} effect. From our conventional analysis (analysis A) a ground state feeding of 15.01 \% is obtained, the optimized branching ratio matrix analysis result is larger, amounting to 20.23 \% , but still in agreement with the ENSDF value within the error interval (15 (8) \%). The ground state value of the optimized branching ratio matrix analysis agrees better with the recently published preliminary results of  Fija\l{}kowska {\it et al.} \cite{Fijalkowska} that also use the total absorption technique, which show a value above 20 \%. 
Our analyses provide no feeding to levels at 2250 keV ($4^{+}$) and at 2350 keV ($2^{+}$), also pointing to the possibility that the  {\it Pandemonium} effect affects these levels, when compared with the high resolution results. 

In Table \ref{HeatBr} we present a comparison of the deduced mean energies from the present work with the values obtained from high resolution measurements. As in the $^{91}$Rb case, we provide the value obtained from the optimized branching ratio analysis result. The value obtained for the electromagnetic component is 358 keV smaller than the preliminary values obtained by Fija\l{}kowska {\it et al.} \cite{Fijalkowska} (4110 (411) keV) determined with a large uncertainty. In this last publication \cite{Fijalkowska} no details of the specific assumptions for the analysis of this decay were given, so we can not discuss further the possible sources of difference with the results of our analysis. 

\begin{table}[h]	
\caption{\label{HeatBr}Mean average energy for $\beta$-particles and $\gamma$ rays (all collected photons) from the decay of $^{86}$Br compared with the value included in the ENSDF database.}
\begin{ruledtabular}
\begin{tabular}{|l|c|c|}
&$\bar{E_\gamma}$ [keV]&$\bar{E_\beta}$ [keV]\\\hline
Present result&3782(116)&1687(60)\\
ENSDF & 3666(109) & 1875(295) \\
\end{tabular}
\end{ruledtabular}
\end{table}

\section*{Discussion and conclusions}

This work has presented the study of the beta decay of $^{86}$Br and $^{91}$Rb using the total absorption technique. Both decays are considered to be important contributors to the decay heat in reactors \cite{Nichols,Yoshida,Gupta2010} and were shown to suffer from the {\it Pandemonium} effect. The decays were studied using isotopically pure beams provided by the IGISOL facility using the JYFL Penning trap and a recently developed total absorption detector. 
The decay of $^{91}$Rb is of particular interest, because this decay was used as a normalization point in the systematic studies of Rudstam {\it et al.} \cite{Rudstam90}, where it was assumed that this decay does not suffer from the {\it Pandemonium} effect. This decay was also measured by Greenwood {\it et al.}, \cite{Greenwood} so it is possible to compare both TAGS results and establish possible systematic differences arising from the different analysis techniques used. On the one hand our present results for $^{91}$Rb agree quite well with the results of Greenwood {\it et al.} On the other hand the deduced mean gamma energy associated with this decay differs from the high resolution value used by Rudstam {\it et al.} pointing to the necessity of renormalizing the gamma energies of that work. 

\begin{figure}[ht]
 \begin{center}
 \includegraphics[width=8.6cm]{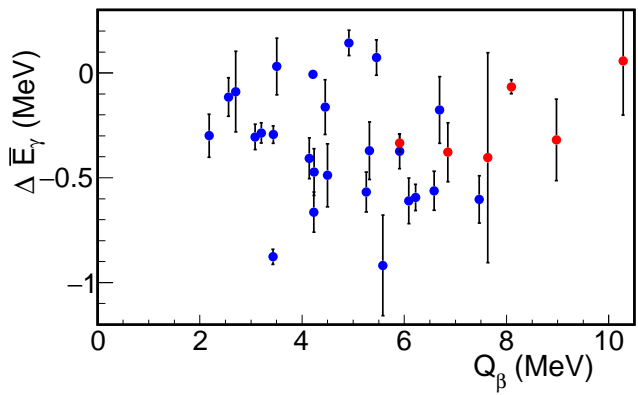}
 \caption{\label{TAGS_Rudstam_1} (Color online) Differences between the mean energies reported in the work of Rudstam {\it et al.} \cite{Rudstam90} and the deduced mean gamma energies from the work of Greenwood {\it et al.} \cite{Greenwood} and our recent data (in red) \cite{Valencia,Tain15,Zak15}. }
\end{center}
\end{figure}

\begin{figure}[ht]
 \begin{center}
 \includegraphics[width=8.6cm]{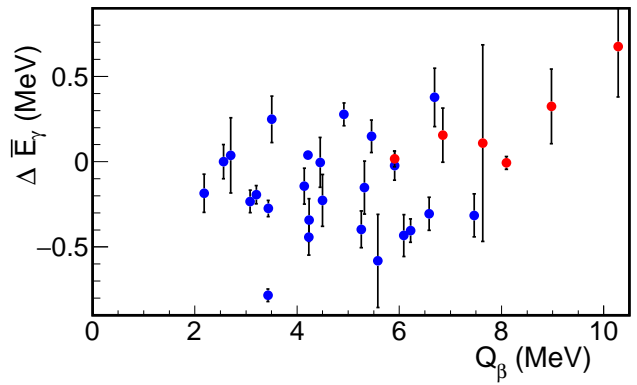}
 \caption{\label{TAGS_Rudstam_2} (Color online) Same as Fig. \ref{TAGS_Rudstam_1} but renormalizing the mean energies reported in Rudstam {\it et al.} \cite{Rudstam90}  by the 1.14 value deduced in this work.} 
\end{center}
\end{figure}

It was pointed out by O. Bersillon in one of the earlier meetings of the WPEC25 \cite{Nichols,Yoshida}, that there are large discrepancies between the mean energies deduced from the TAGS results of Greenwood and the Rudstam results. In particular,  the Rudstam mean gamma energies are systematically smaller than the corresponding mean energies deduced from the Greenwood TAGS data. One might think that the source of the discrepancy lies in the incorrect normalization value. So, this is an issue that can be revisited using the new normalization of the Rudstam data set presented in this article. In the comparison presented here we have also included the mean energies deduced for some cases of our recent TAGS work for which the differences with Rudstam data can be calculated ($^{86,87,88}$Br, $^{91,92,94}$Rb \cite{Valencia,Tain15,Zak15}). The comparison is presented in Fig. \ref{TAGS_Rudstam_1} first using the original Rudstam results and then in Fig. \ref{TAGS_Rudstam_2} using the renormalized results of Rudstam with our present value of the mean gamma energy of the $^{91}$Rb decay. The results show that even though the relative differences are reduced, there is a remaining systematic difference between the two sets of results. The mean value of the differences in the mean gamma energies changes from -360 keV to -180 keV after the renormalization by 1.14. In any case the most striking fact is the large spread of the observed differences ranging from $-0.8$~MeV to $+0.6$~MeV
even after the normalization. There seems to be no systematic trend. At present the origin of such discrepancies is not clear.

\begin{figure}[ht]
 \begin{center}
 \includegraphics[width=8.6cm]{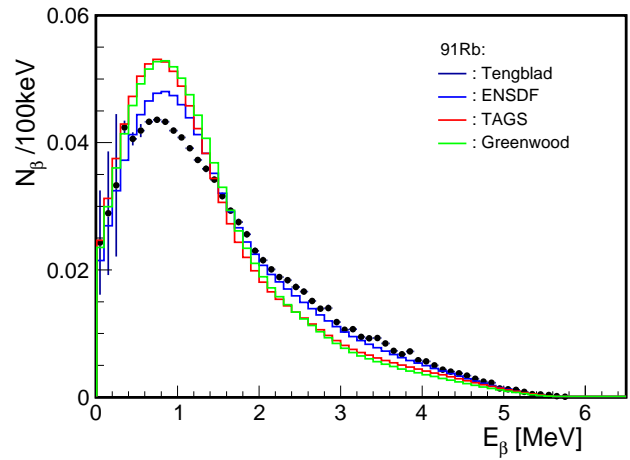}
 \caption{\label{Rb91_beta} (Color online) Comparison of the beta spectrum deduced from our TAGS measurements, Greenwood measurements and from ENSDF, assuming allowed transitions, with the measurements of Tengblad {\it et al.} \cite{Tengblad}.}
\end{center}
\end{figure}

\begin{figure}[ht]
 \begin{center}
 \includegraphics[width=8.6cm]{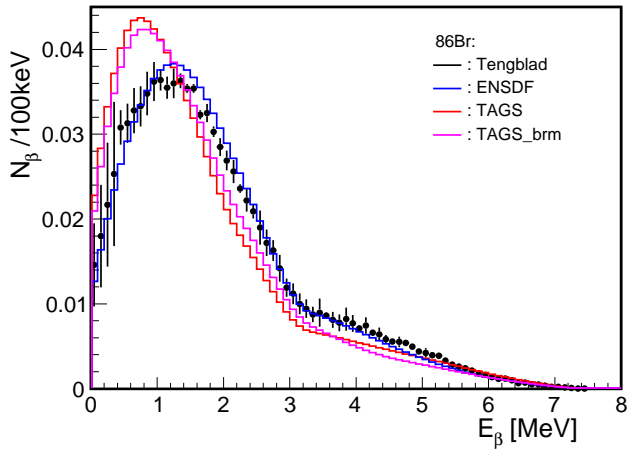}
 \caption{\label{Br86_beta} (Color online) Comparison of the beta spectrum deduced from our TAGS measurements for both analyses presented in this work, and from ENSDF, assuming allowed transitions, with the measurements of Tengblad {\it et al.} \cite{Tengblad}.}
\end{center}
\end{figure}

It is also possible to deduce the beta spectrum from the TAGS data for both measured cases and compare them with the measurements of Tengblad {\it et al.} \cite{Tengblad,Rudstam90}. This comparison is also relevant because one of the cross-checks employed in Rudstam's publication is the comparison of the sum of the mean gamma, beta and deduced antineutrino mean energies with the Q value of the decay. If there is a systematic difference in the mean gamma energies, we can expect possible systematic differences also in the beta decay energies and in the deduced beta spectra.  This is presented in Fig. \ref{Rb91_beta} for $^{91}$Rb  decay and in Fig. \ref{Br86_beta} for the $^{86}$Br decay. The beta spectrum has been deduced assuming allowed shape transitions and using the subroutines of the program LOGFT of the NNDC (Brookhaven) \cite{NNDC}. We see systematic differences in the beta spectrum of both decays. These differences can not be explained by the assumption of the allowed character of the beta transitions used in the deduction of the spectra from the TAGS measurements. Actually if we assume first forbidden transitions (using the procedure employed in the LOGFT utility of NNDC) for all beta transitions the deduced beta spectrum does not differ so much from the one obtained assuming allowed transitions and presented here \cite{AlgoraND2016}. For the present cases and for the recently studied $^{87,88}$Br and $^{94}$Rb cases \cite{Valencia} we can see that the deduced beta spectrum from TAGS measurements is systematically softer (shifted to lower energies) than the directly measured Tengblad data \cite{Tengblad}. This can be an important issue to be taken into account for antineutrino summation calculations using different data sets. 

\begin{figure}[ht]
 \begin{center}
 \includegraphics[width=8.6cm]{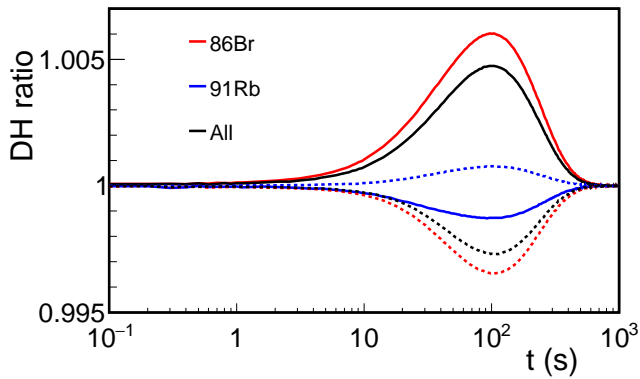}
 \caption{\label{decay_heat235} (Color online) Impact of the new TAS data relative to the high resolution data on the decay heat of $^{235}$U.
The continuous line represents the electromagnetic component, the dotted line the light particle component.}
\end{center}
\end{figure}

\begin{figure}[ht]
 \begin{center}
 \includegraphics[width=8.6cm]{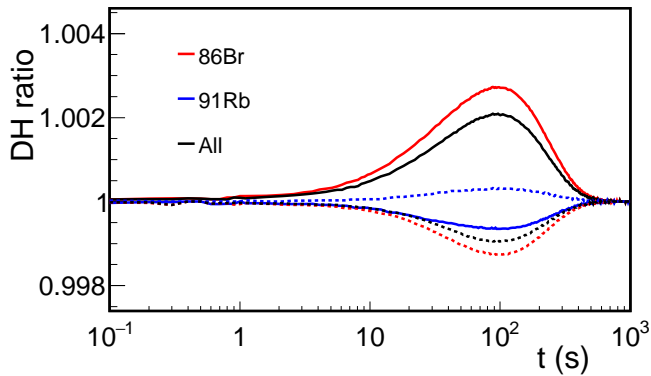}
 \caption{\label{decay_heat239} (Color online) Impact of the new TAS data relative to the high resolution data on the decay heat of $^{239}$Pu
(for details see Fig. \ref{decay_heat235}).}
\end{center}
\end{figure}

\begin{figure}[ht]
 \begin{center}
 \includegraphics[width=8.6cm]{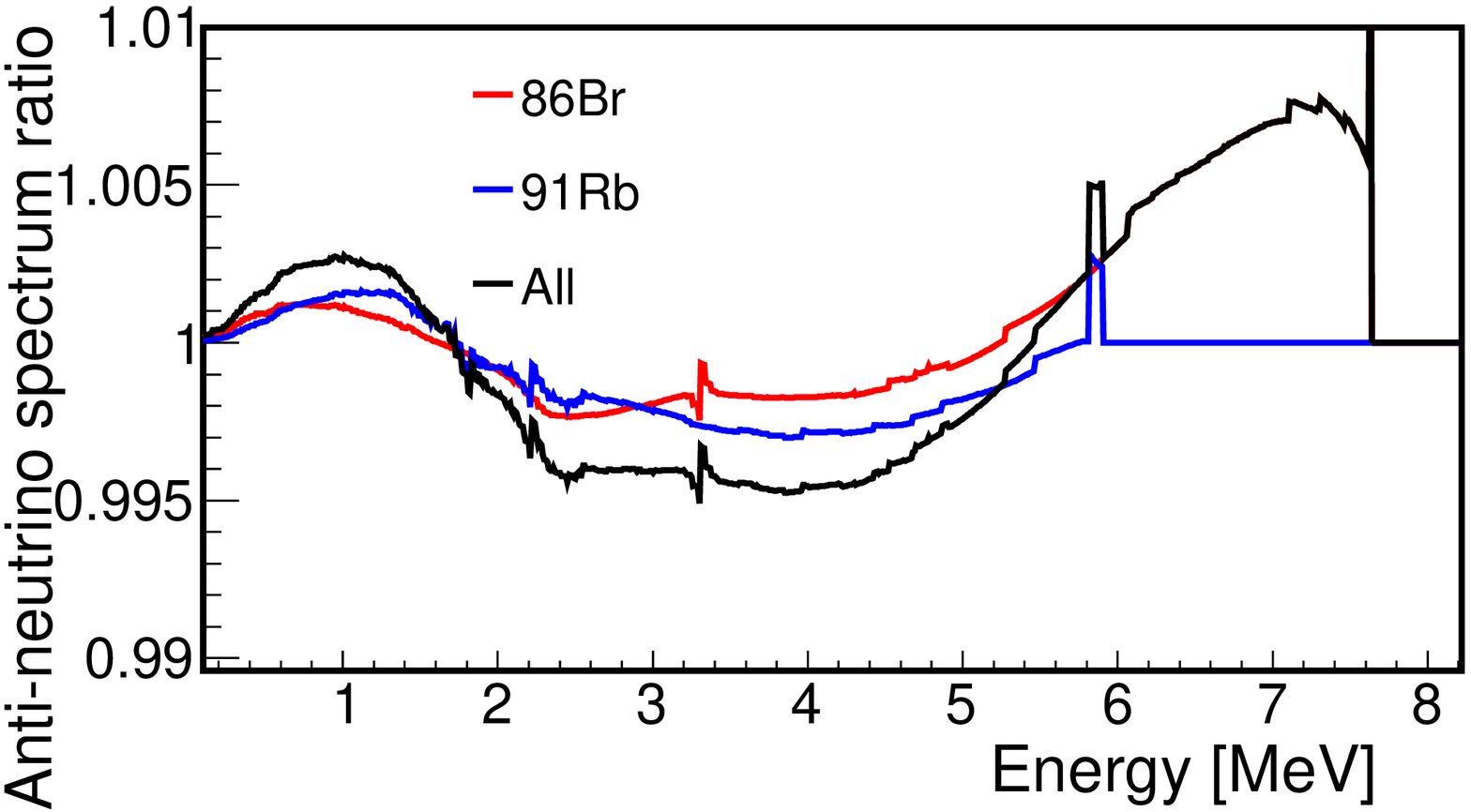}
 \caption{\label{nu_235} (Color online) Impact of the new TAS data relative to the high resolution data on the antineutrino spectrum of $^{235}$U.}
\end{center}
\end{figure}

\begin{figure}[ht]
 \begin{center}
 \includegraphics[width=8.6cm]{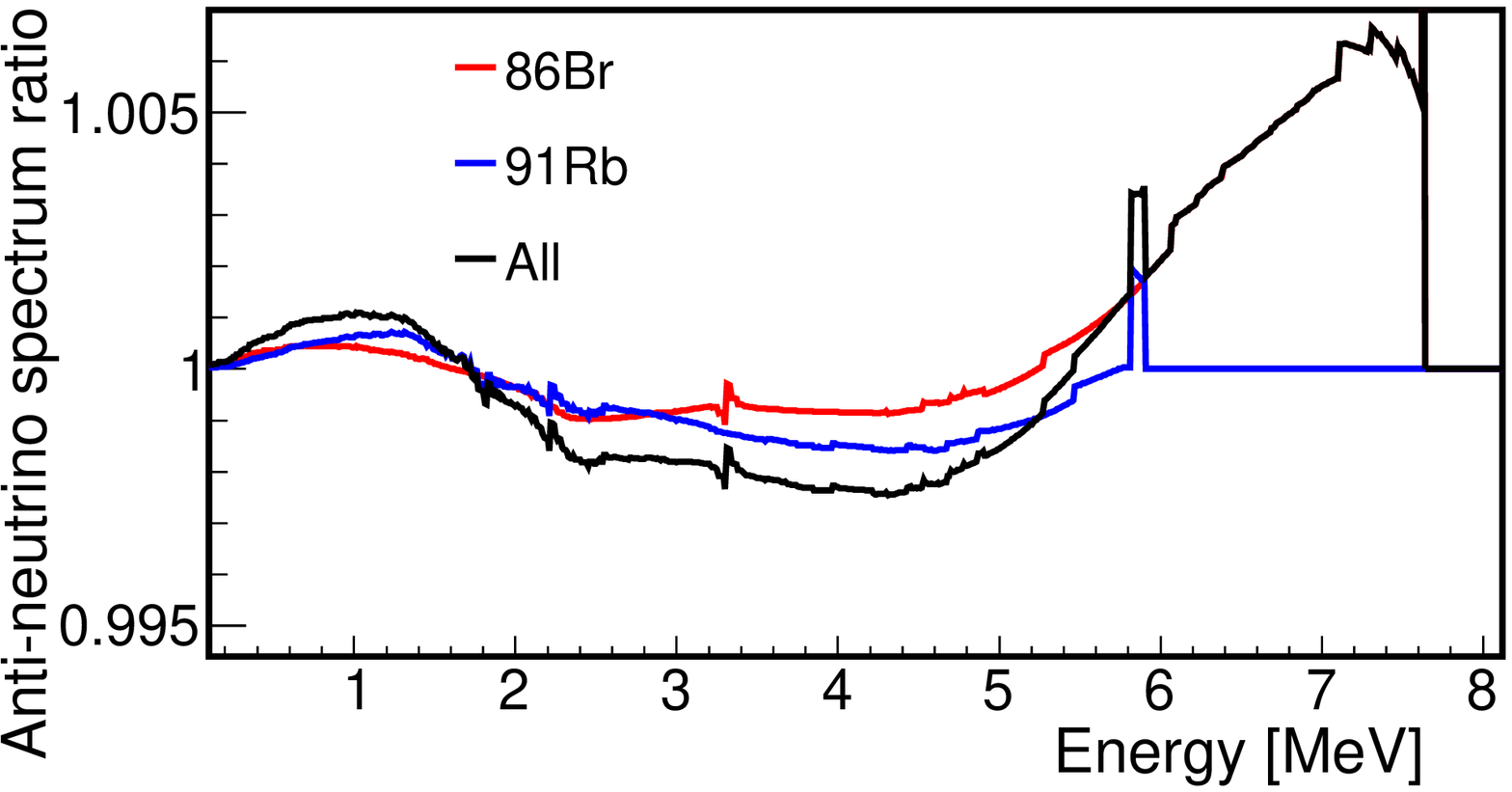}
 \caption{\label{nu_239} (Color online) Impact of the new TAS data relative to the high resolution data on the antineutrino spectrum of $^{239}$Pu.}
\end{center}
\end{figure}

The relative impact of the TAGS data of both decays on the calculations of the decay heat and on the predictions of the antineutrino spectrum is compared in Figs. \ref{decay_heat235}, \ref{decay_heat239} and Figs. \ref{nu_235}, and \ref{nu_239} with respect to high resolution data. 
 They have a small impact on the decay heat calculations and it is more relevant for $^{235}$U than for $^{239}$Pu. As can be seen in Fig. \ref{decay_heat235} it amounts to up to 0.5 \% in $^{235}$U and up to 0.2 \% in $^{239}$Pu for the electromagnetic component. The relative contribution to the light particle component is approximately 0.2 \% for 
 $^{235}$U and 0.1 \% for $^{239}$Pu at its maximum. As in the case of the decay heat, the relative impact on the antineutrino spectrum is more relevant for $^{235}$U and for all fuels ($^{235}$U, $^{238}$U, $^{239}$Pu, $^{241}$Pu) it has the largest contributions at approximately 4 and 7 MeV antineutrino energies, but in opposite directions. At around 3-4 MeV the contribution to the global antineutrino spectrum is reduced in all fuels. At higher energies (above 6 MeV) the contribution is larger and positive and it comes only from the decay of $^{86}$Br that has a larger decay Q value. This latter impact is due to the change in the ground state feeding and affects a region which has partial overlap with the anomaly seen in the antineutrino spectrum centred around 5 MeV \cite{Choi}. But it must be mentioned that the relative impact of this decay is modest. 

In a similar fashion to the antineutrino calculations performed in \cite{Zak15} the maximum impact of the contributions of $^{86}$Br and $^{91}$Rb have been estimated for the antineutrino spectrum from a pressaurised water reactor (PWR) with fuel at equilibrium (52 \% $^{235}$U + 33 \% $^{239}$Pu + 8.7 \% $^{238}$U + 6 \% $^{241}$Pu). In this framework $^{86}$Br has a maximum impact of  1.18\% (1.5\% in $^{235}$U, 1\% in $^{239}$Pu) in the bin 6-7 MeV and 1.04 \% in the bin 5-6 MeV). $^{91}$Rb has a maximum impact of 0.99\% in the bin 4-5 MeV (1.3\% in $^{235}$U). These estimates were obtained using the original Rudstam spectra in the summation calculations. The two nuclei have a moderate impact on the antineutrino spectra, as foreseen since they contribute at most 1.5\% to the $^{235}$U  antineutrino energy spectrum. Nevertheless, provided that Rudstam {\it et al.} measured spectroscopic information for 111 nuclei, the impact on antineutrino spectra built with the summation method of an eventual systematic bias affecting these spectroscopic data may be larger and has to be assessed. 


In the introduction it is emphasised that the decays studied are of high relevance for nuclear applications. It appears, however, from our results that the impact on both decay heat and the reactor antineutrino spectrum is relatively modest. The reader should note that it is only the relative impact of the new TAGS results compared with the high resolution studies that is modest. Both decays are important contributors to the decay heat in the cooling time range of 100 s, as can be seen in the reactor decay heat calculations presented by  Fleming and Sublet in \cite{CCFE_report}. The contributions of the $^{86}$Br and $^{91}$Rb decays can amount up to 3.9 \% and 8.9 \% respectively in the gamma component of the decay heat in $^{235}$U and up to 1.7 \% and 4.2\% respectively in $^{239}$Pu.

This work was supported by Spanish Ministerio de Econom\'{\i}a y
Competitividad under grants FPA2008-06419, FPA2010-17142 and FPA2011-24553 and 
FPA2014-52823-C2-1-P, FIS2014-51971-P, CPAN CSD-2007-00042 (Ingenio2010), and
the program Severo Ochoa (SEV-2014-0398), by EPSRC and STFC (UK) and by the CHANDA
European project, the In2p3 institute of CNRS, and the NEEDS challenge through the NACRE project.
Work at ANL was supported by the U.S Department of Energy under contract 
DE-AC02-06CH11357. The authors would like to thank the late Olivier Bersillon for drawing 
our attention to the discrepancy between the mean energies deduced from the Greenwood TAGS data 
and the Rudstam data. The support of working groups of the IAEA and the IAEA CRP in the identification of priority nuclei for the decay heat calculations is acknowledged.


\begin{thebibliography}{00}



\bibitem{alg10} A. Algora {\it et al.}, Phys. Rev. Lett. {\bf 105}, 202501 (2010).

\bibitem{Fallot} M. Fallot {\it et al.}, Phys. Rev. Lett. {\bf 109}, 202504 (2012).

\bibitem{Rudstam90} G. Rudstam {\it et al.}, Atomic Data and Nuclear Data Tables {\bf 45}, 239 (1990).

\bibitem{Tengblad} O. Tengblad {\it et al.}, Nucl. Phys. A 503, 136 (1989).

\bibitem{ensdf} ENSDF, http://www.nndc.bnl.gov/ensdf.

\bibitem{har77} J. Hardy {\it et al.}, Phys. Lett. B {\bf 71}, 307 (1977).

\bibitem{alg99} A. Algora {\it et al.}, Nucl. Phys. A {\bf 654},  727c (1999).

\bibitem{hu99} Z. Hu {\it et al.}, Phys. Rev. C {\bf 60}, 024315 (1999).

\bibitem{alg03} A. Algora {\it et al.}, Phys. Rev. C 68, 034301 (2003). 

\bibitem{Nichols}  A. L. Nichols, {\it et al.},  Beta decay and decay heat, INDC(NDS)-0499 (2006).

\bibitem{Yoshida} T. Yoshida {\it et al.}, Assessment of Fission Product Decay Data for Decay Heat Calculations, OECD/NEA Working Party for International Evaluation Co-operation, Volume 1425 25, 2007.

\bibitem{Gupta2010} M. Gupta {\it et al.}, Decay Heat Calculations: Assessment of Fission Product Decay Data Requirements for Th/U Fuel, IAEA report INDC(NDS)-0577, 2010.

\bibitem{Zak15} A.-A. Zakari-Issoufou {\it et al.}, Phys. Rev. Lett.  {\bf 115}, 102503 (2015).

\bibitem{Tain15} J. L. Tain {\it et al.},  Phys. Rev. Lett.  {\bf 115}, 062502 (2015).

\bibitem{Wang12} M. Wang {\it et al.},  Chinese Physics C 36, 1603 (2012).

\bibitem{Greenwood} R. C. Greenwood, R. G. Helmer, M. H. Putnam and K. D. Watts, Nucl. Instrum. Methods Phys. Res. A, {\bf 390}, 95 (1997).

\bibitem{Guz10} R. Rodriguez-Guzman, P. Sarriguren, and L. M. Robledo, Phys. Rev. C 82, 061302(R) (2010).

\bibitem{Nacher} E. N\'acher {\it et al.}, Phys. Rev. Lett. 92, 232501 (2004).

\bibitem{Poirier} E. Poirier {\it et al.}, Phys. Rev. C 69, 034307 (2004).

\bibitem{Perez} A. P\'erez-Cerdan {\it et al.}, Phys. Rev. C 84, 054311 (2011).

\bibitem{Briz} J. A. Briz {\it et al.}, Phys. Rev. C92, 054326 (2015).

\bibitem{Aguado} M. E. Est\'evez Aguado {\it et al.}, Phys. Rev.  C 92, 044321 (2015).

\bibitem{Scheck16} M. Scheck  {\it et al.}, Phys. Rev. Lett. 116, 132501 (2016). 

\bibitem{ays01} J. \"{A}yst\"{o}, Nucl. Phys. A {\bf 693}, 477 (2001).

\bibitem{kol04} V. Kolhinen {\it et al.}, Nucl. Instrum. Methods Phys. Res. A {\bf 528}, 776 (2004).

\bibitem{ero12} T. Eronen {\it et al.}, Eur. Phys. J. A {\bf 48}, 46 (2012).

\bibitem{cano1999pulse} D. Cano-Ott {\it et al.}, Nucl. Instrum. Methods Phys. Res. A  {\bf 430}, 488 (1999).

\bibitem{tai07a} J. L. Tain {\it et al.}, Nucl. Instrum. Methods Phys. Res. A {\bf 571}, 719 (2007).

\bibitem{tai07b} J. L. Tain {\it et al.}, Nucl. Instrum. Methods Phys. Res. A {\bf 571}, 728 (2007).

\bibitem{Capote2009} RIPL-3, R. Capote {\it et al.}, Nucl. Data Sheets {\bf 110}, 3107 (2009).

\bibitem{Gilbert} A. Gilbert and A. G. W. Cameron. Canadian Journal of Physics {\bf 43} 1446 (1965).

\bibitem{BSFG_Dilg} W. Dilg, W. Schantl, H. Vonach, and M. Uhl. Nuclear Physics A {\bf 217} 269 (1973).

\bibitem{CT} T. Von Egidy, H.H. Schmidt, and A.N. Behkami.  Nuclear Physics A  {\bf 481} 189 (1988).

\bibitem{Goriely2001} S. Goriely, F. Tondeur, and J. Pearson, At. Data Nucl. Data Tables  {\bf 77}, 311 (2001).

\bibitem{Demetriou2001} P. Demetriou and S. Goriely, Nucl. Phys. A {\bf 695}, 95 (2001).

\bibitem{cano1999monte} D. Cano-Ott {\it et al.}, Nucl. Instrum. Methods Phys. Res.,
Sect. A {\bf 430}, 333 (1999).

\bibitem{agostinelli2003} S. Agostinelli {\it et al.}, Nucl. Instrum. Methods Phys. Res. A {\bf 506}, 250 (2003). 	 
\bibitem{Baglin13} C. M. Baglin, Nuclear Data Sheets {\bf 114}, 1293 (2013).

\bibitem{Greenwood2} R. C. Greenwood, M. H. Putnam and K. D. Watts, Nucl. Instrum. Methods Phys. Res. A, {\bf 378}, 312 (1996).

\bibitem{Sarriguren} P. Sarriguren, E. Moya de Guerra, and A. Escuderos, Phys. Rev. C 64, 064306 (2001); P. Sarriguren, Phys. Rev. C 79, 044315 (2009).

\bibitem{ensdf86} A. Negret and B. Singh, Nuclear Data Sheets {\bf 124}, 1 (2015).

\bibitem{Fotiades} N. Fotiades, M. Devlin, R. O. Nelson, and T. Granier, Phys. Rev. C {\bf 87}, 044336 (2013).

\bibitem{ensdf86_old} B. Singh, Nuclear Data Sheets {\bf 94}, 1 (2001).

\bibitem{Porquet09} M.-G. Porquet, {\it et al.}, Eur. Phys. Journal A {\bf 40}, 131 (2009).

\bibitem{Fijalkowska} A. Fijalkowska {\it et al.}, Acta Phys. Polonica B {\bf 45}, 545 (2014).

\bibitem{Valencia} E. Valencia {\it et al.}, Phys. Rev. C 95, 024320 (2017).

\bibitem{NNDC} ENSDF Analysis Programs - LOGFT, National Nuclear Data Center, Brookhaven National Laboratory, http://www.nndc.bnl.gov/nndcscr/ensdf pgm/analysis/logft/.

\bibitem{AlgoraND2016} A. Algora {\it et al.}, contribution to the ND2016 conference, in print.

\bibitem{Choi} J. H. Choi {\it et al.},  Phys. Rev. Letts. 116, 211801 (2016).




\bibitem{CCFE_report} M. Fleming and J.-C. Sublet, CCFE-R(15)28/S1 Report, June 2015. 





\end{thebibliography}
\end{document}